# A Note on the Rapanui Lunar Calendar

Sergei Rjabchikov[1]

[1]The Sergei Rjabchikov Foundation - Research Centre for Studies of Ancient Civilisations and Cultures, Krasnodar, Russia, e-mail: srjabchikov@hotmail.com

## Abstract

This paper is dedicated to the research of secrets of Easter Island (Rapa Nui), a remote plot of land in the Pacific; the article contains not only ethnological data, but also some results on the archaeoastronomy. This author examines both lunar calendar lists presented on the Mamari tablet as well as on a panel at Ahu Raai. Calendar records on different tablets have been investigated, too.

## Data on the Nights/Days of the Month

The author uses the own methodology of the decipherment of the local hieroglyphic script known *as rongorongo*.[2] Different calendar records have already been decoded.[3]

Here and everywhere else, I use the computer program RedShift Multimedia Astronomy (Maris Multimedia, San Rafael, USA) to look at the heavens above Easter Island.

The natives counted 29 and 30 moons (lunar phases) in two consecutive months, so the average duration of each month lasted 29.5 nights/days. The names of nights/days of moon age were as follows:

1. *Hiro* (the new moon; it was invisible; *Whiro* in the Maori calendar of New Zealand)
2. *Tireo, Tueo* (this moon was invisible; *Tirea* in the Maori calendar)
3. *Ata*
4. *Ari*
5. *Kokore tahi*
6. *Kokore rua*
7. *Kokore toru*
8. *Kokore ha*
9. *Kokore rima*
10. *Kokore ono*
11. *Maharu*
12. *Hua*
13. *Atua* (the almost full moon)
14. *Hotu* (the full moon)
15. *Ma-ure* (the full moon)
16. *(Ina-)ira* (the full moon)
17. *Rakau* (the full moon)
18. *Ma-tohi* (the "Bearing" moon; the end of the list of the quasi-full phases of the moon)
19. *Kokore tahi*

20. *Kokore rua*
21. *Kokore toru*
22. *Kokore ha*
23. *Kokore rima*
24. *Tapu mea* (= *Tangaroa* in other Polynesian calendars)
25. *Matua*
26. *Rongo*
27. (*Rongo*) *Tane*
28. *Mauri-nui*
29. *Mauri-kero* (this moon was invisible)
(30. *Mutu*) (this moon was invisible).

**The Vast Calendar Record on the Mamari Tablet**

Consider the following text on the Mamari tablet (C), see figure 1.

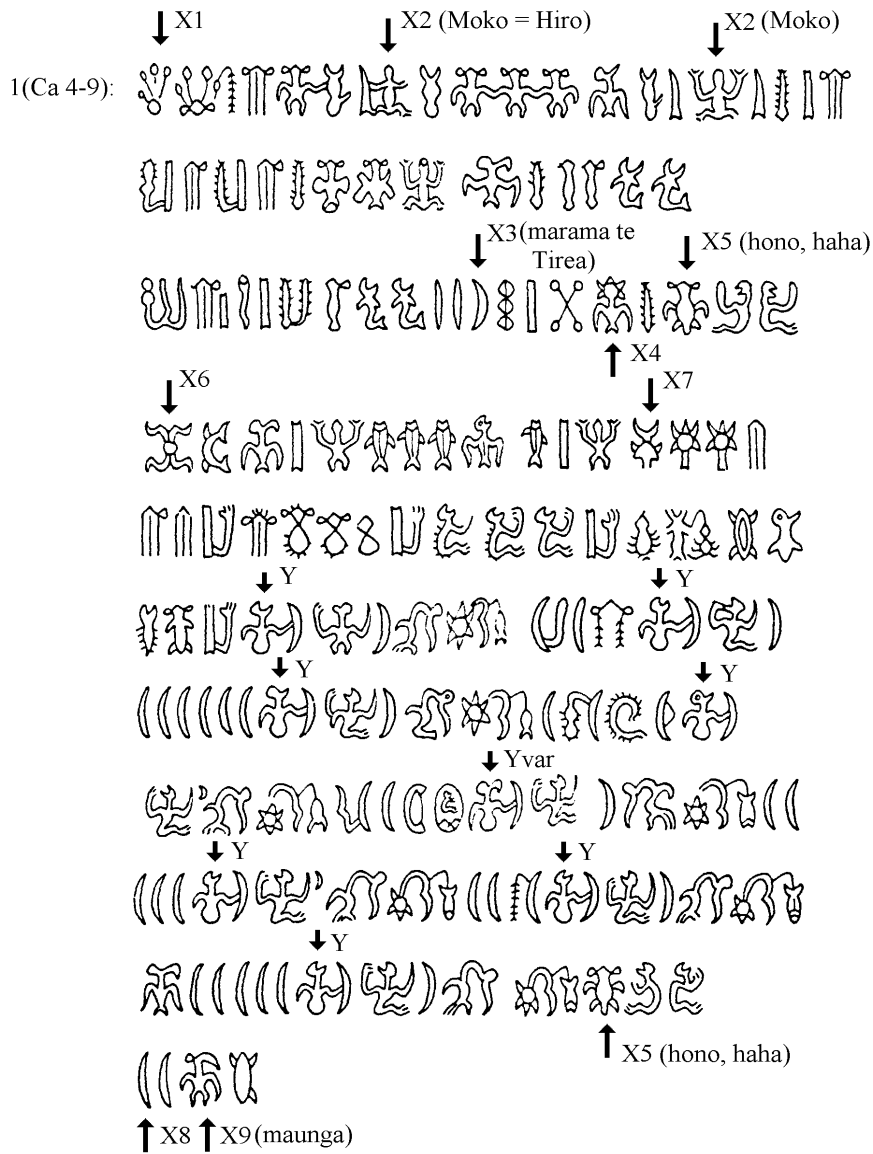

Figure 1.

This inscription deals with the summer solstice that occurred on the second day (*Tireo*) of the lunar calendar. It should be borne in mind that in the Old Rapanui calendar the night/day *Hiro* was the first, it was a day of the possible solar eclipse; the night/day *Tirea* (*Tireo*, *Tueo*) was the second. Perhaps, the text begins with the report concerning the new moon of December 19, A.D. 1672:

Fragment X1 reads **137a 137b 24 26** *Raa TEA raa ari, maa* '(It was) the bright sun;'

Fragments X2 read **69** *Moko* (The Lizard, the night/day *Hiro*);

Fragment X3 reads **3 17 4-40** *marama te Tirea* (*Tireo*, *Tueo*, Maori *Tirea*) 'the moon *Tireo*;'

Fragment X4 reads **44 7 25** *Ta(h)a Tuu Hua* 'The star Aldebaran (α Tauri) turned (in the sky);'

Fragment X5 reads **68 6-6** *hono, haha* 'add, gaze (at those two invisible crescents)!;'

Fragment X6 reads **108 8** *Hiri Vaka* 'The stars β and α Centauri [*Nga Vaka*] rose (in the sky);'

Fragment X7 reads **21 3 7-7 26-26-26 4 15** *ko marama tuutuu maa, maa, maa atua roa* '(It was) the great deity of the come month of the summer solstice;'

Fragments Y read **68 3 6 3 19 7 50 12** (or **16**) *Hono marama a marama, ku tuu i Ika* (or *Kahi*) 'A moon was joined (added) to another moon. (The moon) came to the Fish = *Hina*, the first part of the month (or to the Tuna Fish = *Tangaroa*, the second part of the month);' notice that in these fragments the crescents accompanied with some names appear;

Fragments X5-X8-X9 read **68 6-6 3 3 49-28** *Hono, haha marama, marama maunga* '(So, it was) added gazing at the two crescent after the end (of the lunar month).'

Thus, the count of phases of the moon is marked by turtle glyphs **68** *hono*, *honu*, cf. Rapanui *hono* 'to add.'

## The Parallel Calendar Record at Ahu Raai

In this connection one can study a Rapanui rock design on a panel at the ceremonial platform Ahu Raai,[4] see figure 2.

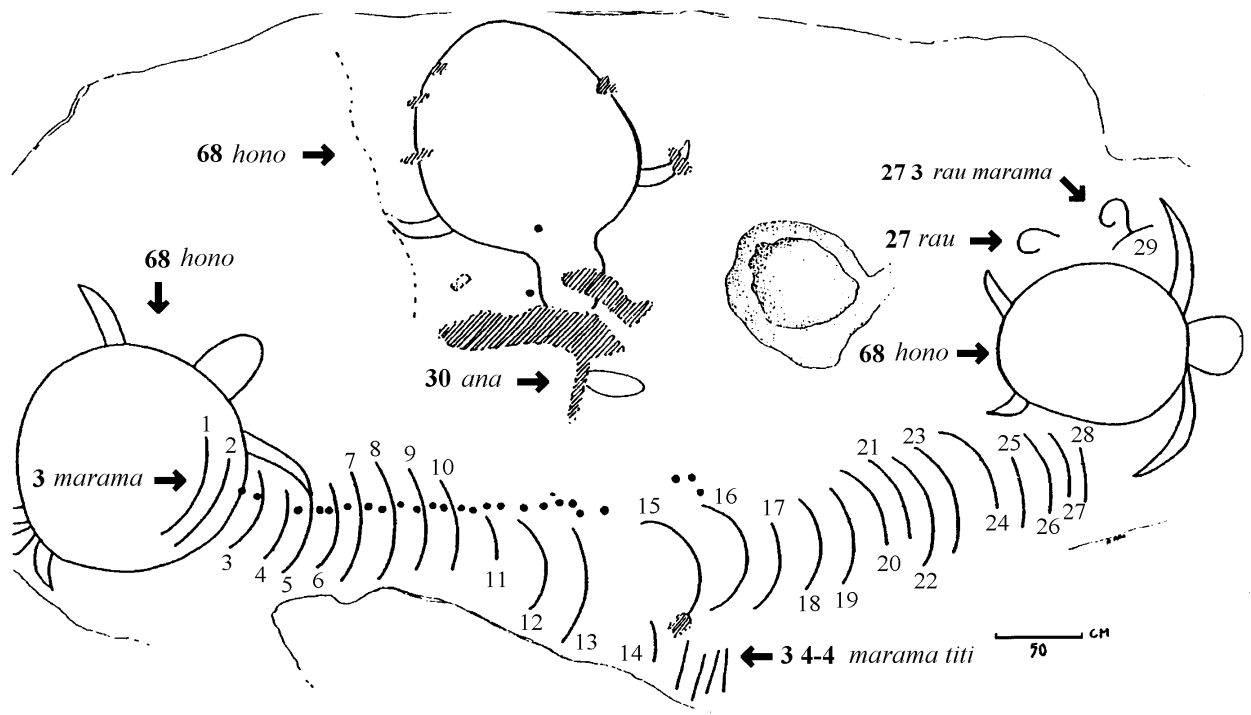

Figure 2.

In accordance with Lee, 28 curved lines are relevant to a moon count. I prefer to distinguish such 29 crescent lines, they have numbers from 1 to 29 in this figure (the last one is depicted in the conjunction with the hook glyph above the third turtle glyph). Curved lines in different Polynesian rock designs correspond to glyphs **3** *marama* (*hina*) 'moon.'[5]

The three turtle glyphs **68** *hono*, *honu* 'to add' are markers of the count. Glyph **30** *ana* with the meanings 'abundance; many times; too much' is discovered in local rock designs together with drawings of sea creatures, cf. Hawaiian *ana* 'to have enough or too much.'[6]

The 14th lunar phase is written down as the glyph combination **3 4-4** (or **4-5**) *marama titi* 'the full moon.' At the end of the rock calendar glyph **27** *rau* is repeated twice, cf. Tongan *lau* 'to count,' and Hawaiian *lau* 'to be numerous or many.' This glyph without the association with the lunar glyph denotes the 30th moon that might be vacant.

The lunar calendar on the Mamari tablet as well as the lunar calendar as a rock design at Raai have been compared earlier.[7] I should like to stress that the board was manufactured in the eastern part of the island (the Tupahotu tribe) in the 17th century A.D.,[8] and the rock drawing was made on the same territory (the same tribe) as well. It is the reliable clue to the dating of this masterpiece of the local art.

### Additional Parallels on Two Tablets

Consider two parallel calendar records which were taken down on the Aruku-Kurenga (B) and Tahua (A) tablets, see figure 3.

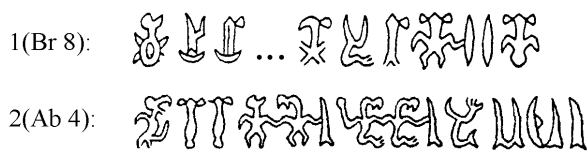

1(Br 8):

2(Ab 4):

Figure 3.

1 (Br 8): **6-28 3 5-5 3 26 … 49 2 56 6 30-30 68** *Anga Hina titi, Hina maa … Mau Hina po, ha anaana, hono.* 'The full moon (the bright moon) is moving… The moon goddess *Hina* THE NIGHT is abundant, (the moons) have been added, (it is) the count (of lunar phases).'

Old Rapanui *anga* means 'to move in a certain direction,' cf. Maori *anga* 'ditto.'

2 (Ab 4): **19 56-56 6 62-5 62-62-5 2 5-5 3 5-5** *Ku popo; ha toti, tototi Hina titi, hina titi.* 'The moon became full; the moon goddess *Hina* who is full and the full moon join (the month).'

Old Rapanui *toti* 'attached' is registered in the name *Humu-toti* in the Rapanui mythology,[9] cf. Tahitian *toti* 'attached.'

### The Names of Phases of the Moon

Consider the following records with the names of some lunar phases on the Great Santiago (H), Small Santiago (G), Aruku-Kurenga (B), Great Washington (S), Keiti (E), Tahua (A) and Mamari (C) tablets, see figure 4.

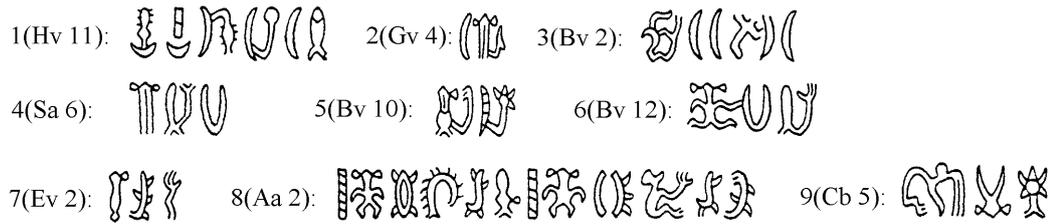

Figure 4.

1 (Hv 11): **3 25 3 4 3 25 3 4 3 12** *Marama Hua, marama Atua, marama Hua, marama Atua, marama ika.* 'The lunar phase *Hua*, the lunar phase *Atua*, the lunar phase *Hua*, the lunar phase *Atua*, the lunar phases of (the successful) fishery.'

It was an exercise for pupils in the royal *rongorongo* school. So, the names of the nights/days *Hua* and *Atua* (with the bright moon in the sky) were repeated twice. I suppose that both lunar phases marked days when the catch of fish was successful. Really, the almost full or entirely full moon could be a symbol of abundance, plenty and fertility. In this connection let us examine another local rock design at Ahu Raai.[10] Here two canoes are depicted nearby. Above them there are two glyphs: **3 4** *Marama Atua.* 'The moon (lunar phase) *Atua*.' Hence, the night/day *Atua* was good for the fishery indeed.

2 (Gv 4): **3 26-102** *Marama Ma-ure* '(It was) the lunar phase *Ma-ure*.'

3 (Bv 2): **3 50-15 3 3 70 140** *Hina-ira, marama, marama Pua* THE FULL MOON. '(They were) the lunar phase *Ina-ira*, (another) moon (= *Rakau*), (and) the "Bearing" moon (= *Ma-tohi*).'

Old Rapanui *pua* signifies 'to bear,' cf. Mangaian *pua* 'to come forth,' Hawaiian *pua* 'descendants' and Maori *mapua* (*ma-pua*) 'bearing abundance of fruit.'

4 (Sa 6): **26-48-15 61** *Mauri marama.* '(It was) the lunar phase *Mauri-nui*.'

5 (Bv 10): **26-48-15 3 7** *Mauri marama tuu.* 'The moon *Mauri-nui* was coming.'

6 (Bv 12): **69 61 48-15** *Moko Hina* (*marama*) *Uri* '(It was) the lunar phase 'The Lizard' (= *Hiro*), when the moon was dark.'

7 (Ev 2): **56 51-15** *Po Kero.* '(It was) the night *Mauri-kero*.'

8 (Aa 2): **4-6 28 14 51 73 4-6 57 51 6 51 51-15** *Tuha nga auke. He tuha tara keha, Kekero.* '(It was) the time (when) the seaweeds were available (only). (It was) the time (when) the solar rays were pale, (when was the night) *Mauri-kero.'*

This record in the royal school was written down to remember glyph **51** *ke*. In the first sentence the verbal article *he* was omitted, but in the second sentence this article (glyph **73** *he*) was taken down. The text describes the hunger during the winter months; the seaweeds often were the general food in this case. Old Rapanui *tara* means 'solar rays,' cf. Maori *tara* 'ditto.' Old Rapanui *keha* means 'pale; dim,' cf. Maori *keha* 'ditto.'

9 (Cb 5): **44-32 97-7** *Tau Mutu* '(It was) the time (called) *Mutu*.'

## Conclusions

Valuable information has been gained that the Easter Islanders conducted astronomical observations of the sun, the moon, β and α Centauri, and Aldebaran in pre-European time and later. Some records concerning nights/days of moon age can be read and interpreted.

---

[10] See Lee, G., Op.cit., pp. 178-179, figure 6.15.